\documentstyle[aps,multicol,psfig]{revtex}

\draft

\begin{document}

\title{Non conservative Abelian sandpile model with BTW toppling rule}

\author{Alexei V\'azquez$^{1,2}$}

\address{$^1$ Abdus Salam International Center for Theoretical Physics\\
        Strada Costiera 11, P.O. Box 586, 34100 Trieste, Italy}

\address{$^2$ Department of Theoretical Physics, Havana University\\
        San L\'azaro y L, Havana 10400, Cuba}

\maketitle

\begin{abstract}

A non conservative Abelian sandpile model with BTW toppling rule introduced in
[Tsuchiya and Katori, Phys. Rev. E {\bf 61}, 1183 (2000)] is studied. Using a
scaling analysis of the different energy scales involved in the model and numerical
simulations it is shown that this model belong to a universality class different
from that of previous models considered in the literature.
 
\end{abstract}  


\begin{multicols}{2}\narrowtext

\section{Introduction}

Recently Tsuchiya an Katori \cite{katori00} have introduced a non conservative
Abelian sandpile model with a toppling rule similar to that of the well known Bak,
Tang and Wiesenfeld (BTW) sandpile model \cite{bak87}. The model is defined in a
square lattice of size $L$ and an integer energy profile $z_{ij}$ is considered.
Sites with energy below the threshold $z_c=4\alpha\zeta$ are stable. If the energy
at any given site $(i,j)$ exceeds this threshold the site transfer energy to its four
nearest neighbors following the toppling rule: $z_{ij}\rightarrow z_{ij}-z_c$,
$z_{i\pm1j}\rightarrow z_{i\pm1j}+\zeta$ and $z_{ij\pm1}\rightarrow
z_{ij\pm1}+\zeta$. $\zeta$ is an integer number and $\alpha$ is such that $z_c$ is
also integer. The boundaries are assumed to be open and the system is
perturbed by adding a unit of energy at a site selected at random and letting it
evolve until an equilibrium configuration is reached. 

On each toppling event an amount of energy $\epsilon=4\zeta(\alpha-1)>0$ is
dissipated. For $\alpha=1$ the model is conservative, it is just the BTW model but
with a different scale of toppling and energy addition. In the BTW model
\cite{bak87} the energy added to perturb the system is 1 and on toppling an active
site transfers an energy 1 to each neighbor, they are of the same order. In the
model defined above the energy added is still 1 but the energy transferred on
toppling is $\zeta$. In the limit $\alpha=1$ and $\zeta=1$ the BTW model
\cite{bak87} is recovered while for $\alpha=1$ and $\zeta\gg1$ it is similar to the
BTW model but with a uniform driving. 

In the BTW model ($\alpha=1$ and $\zeta=1$) the avalanches can be decomposed in a
sequence of sub avalanches called waves \cite{ivashkevich94} with well defined
finite size scaling properties.  On the contrary, the distribution of the overall
avalanche size $s$ is better described using a multi-fractal analysis
\cite{tebaldi99}. The break down of the finite size scaling has been recently shown
to be a consequence of the existence of correlations in the sequence of waves
\cite{demenech00}. 

For $\alpha>1$ the model is non conservative but still Abelian \cite{katori00}.  In
the thermodynamic limit $L\rightarrow\infty$, exact calculations by Tsuchiya and
Katori yield the mean avalanche size (including avalanches with size zero) $\langle
T\rangle=\epsilon^{-1}$ \cite{katori00}. Since $\epsilon=4\zeta(\alpha-1)$ they
concluded that in the limit $\alpha\rightarrow1$ $\langle T\rangle$ diverges. 
However, as it is shown in sec.  \ref{sec:scaling} this conclusion is wrong because
$\alpha$ cannot goes to zero in an arbitrary way, in order to satisfy the constraint
that $z_c=4\alpha\zeta$ remains integer. Here it is demonstrated that
$\epsilon\geq1$ and, therefore, $\langle T\rangle\leq1$ for all possible values of
$\zeta$ and $\alpha$. 

The main goal of this work is to investigate the scaling properties of this non
conservative BTW like model in the limit $\alpha\rightarrow1$. The main questions
are related with the existence or not of criticality in the conservative limit
$\alpha\rightarrow1$ and if in this limit one recovers the conventional BTW model
$\alpha=1$. From the analysis of the energy scales involved in the model ( sec.
\ref{sec:scaling}) and numerical simulations (sec. \ref{sec:numerical}) it is
concluded that the model is critical when $\alpha\rightarrow1$ but it does not
belong to the universality class of the BTW model. Its relation with other non
conservative models with BTW like toppling rule introduced in the literature
\cite{manna90,ghaffari97,chessa98} is also discussed (sec. \ref{sec:numerical}).

\section{Scaling laws}
\label{sec:scaling}

In this section some scaling laws are derived based on the energy scales involved in
the model. The main idea of this approach is that the balance between input and
dissipation of energy determine the scaling of some magnitudes with the dissipation
per toppling event, following the general guidelines introduced by Vespiganani {\em
et al} \cite{vespignani97}. For this purpose the avalanches are assumed to be
instantaneous and the analysis is focused in the time scale of the driving field. On
each step one adds 1 unit of energy and measure the toppling activity and the energy
dissipated. On each toppling event an amount of energy $\epsilon=4\zeta(\alpha-1)$
is locally dissipated while an amount $4\zeta$ is transferred to nearest neighbors.
For boundary sites part of the energy is also dissipated through the boundary. 

Let $G(r)$ be the Green function \cite{dhar89}, the average number of toppling
events at a distance $r$ from the site where the energy was added. Close to $r=0$
the effect of local dissipation gives an small contribution and the main energy
scale is given by the transport of the energy from active sites to their nearest
neighbors. On the contrary, far from $r=0$ the effects of the local dissipation
becomes more important. How far will depend on certain correlation length $\xi$,
such that for $r\ll\xi$ transport is more important than local dissipation while for
$r\gg\xi$ the opposite occurs.

Thus, there are two characteristic lengths in this model: the system size $L$ and
the correlation length $\xi$. The analysis developed above is valid in the
thermodynamic limit $L\gg\xi$. In this case the dissipation through the boundary of
the system is negligible in comparison with the energy dissipated on each toppling
event. In such a situation the only way to reach an stationary state is to balance
the input of energy from the driving field with the energy locally dissipated. 
Moreover, since $\xi$ is the only characteristic length $G(r)$ is expected to
satisfy the scaling law
\begin{equation}
G(r)=r^{\eta-d}{\cal F}(r/\xi),
\label{eq:1}
\end{equation}
where $\eta$ is an scaling exponent and $d$ is the spatial dimension.

The amount of energy $\delta E_d(r)$ locally dissipated inside an hyper-circle of
radius $r$ is
\begin{equation}
\delta E_d(r)\propto\epsilon\int_0^rd\rho\rho^{d-1}G(\rho)
\propto \epsilon\xi^\eta f(r/\xi),
\label{eq:2}
\end{equation}
where $f(x)=\int_0^x dy y^{\eta-1}{\cal F}(y)$ and the second proportionality is
obtained using eq. (\ref{eq:1}). On the other hand, the average energy transported
through its boundary $\delta E_t(r)$ is given by
\begin{equation}
\delta E_t(r)\propto\zeta r^{d-1}\frac{dG}{dr}(r)
\propto\zeta\xi^{\eta-2}g(r/\xi),
\label{eq:3}
\end{equation}
where $g(x)=d[x^{\eta-2}{\cal F}(x)]/dx$ and the second proportionality is obtained
using eq. (\ref{eq:1}). The correlation length $\xi$ can be defined as the radius
$r$ at which this two contribution becomes of the same order.  With this definition
and equating eqs. (\ref{eq:2}) and (\ref{eq:3}) with $r=\xi$ it results that
\begin{equation}
\xi\sim\left(\frac{\zeta}{\epsilon}\right)^{\nu},\ \ \ \ \nu=\frac{1}{2}.
\label{eq:4}
\end{equation}

On the other hand, on each step 1 unit of energy is added and in average the amount
$\epsilon\langle T\rangle$ is dissipated, where $\langle T
\rangle\propto\int_0^\infty dr r^{d-1}G(r)$ is the mean avalanche size counting
even the case when it has size 0. Equating this two contributions it results that
\begin{equation}
\langle T\rangle=\frac{1}{\epsilon}=\frac{1}{4\zeta(\alpha-1)}.
\label{eq:5}
\end{equation}
Moreover, using eq. one obtains
(\ref{eq:1})
\begin{equation}
\eta=0.
\label{eq:6}
\end{equation}

Eq. (\ref{eq:5}) reproduces the exact result by Tsuchiya and Katori. The present
approach is however based on more general arguments an can be easily adapted to any
sandpile model with local dissipation.  The same argument (energy balance) has been
previously used by Vespiganani {\em et al} \cite{vespignani97} to understand the
scaling properties of other sandpile models with local dissipation. Here, a slightly
different approach has been considered where the new parameter $\zeta$, the ratio
between the energy received from nearest active neighbors and from the driving
field, has been taken into account.

From eq. (\ref{eq:5}) Tsuchiya and Katori concluded that when $\alpha\rightarrow1$
$\langle T\rangle$ diverges. However, this conclusion is not valid if
$z_c=4\zeta\alpha$ is restricted to be an integer number. To show this let us write
$\alpha=1+\epsilon/4\zeta$ which follows from eq. (\ref{eq:5}). But
$4\zeta\alpha=4\zeta+\epsilon$ is restricted to take integer values. With $\zeta$
being an integer number the only way to satisfy this requirement is that $\epsilon$
is also integer, i.e. $\epsilon=1,2,3,\ldots$. Then, since the smaller non negative
integer is 1 it is concluded that $\epsilon\geq1$ and, therefore, from eq. 
(\ref{eq:5}) $\langle T\rangle\leq1$, i.e. it is bounded. 

Nevertheless, the correlation length $\xi$ in eq. (\ref{eq:4}) does not only depend
on $\epsilon$ but also on $\zeta$. For fixed $\epsilon$ it diverges in the limit
$\zeta\rightarrow\infty$ and the model is critical. The real control parameter is
then $\epsilon_{eff}=\epsilon/\zeta$, i.e. the energy dissipated per toppling event
relative to the characteristic energy scale of transport $\zeta$. Although this
result is in complete agreement with the field theory approach of Vespignani {\em et
al} \cite{vespignani97} the fact that $\langle T\rangle$ does not diverges when
$\epsilon_{eff}\rightarrow0$ ($\zeta\rightarrow\infty$) excludes this model from
their analysis. 

Moreover, in previous sandpile models conservation implies the scaling law $\langle
s\rangle\sim \xi^2$, where $\langle s\rangle$ is the mean avalanche size excluding
those with size 0 \cite{vespignani97}. To investigate the validity of such
scaling relation for the present model let us take into account that $\langle
s\rangle$ is related to $\langle T\rangle$ through the expression
\begin{equation}
\langle s\rangle=\langle T\rangle/P_a,
\label{eq:7}
\end{equation}
where $P_a$ is the probability to obtain an avalanche with non zero size. In the
models considered by Vespignani {\em et al} \cite{vespignani97} $\zeta=1$ and,
therefore, from eqs. (\ref{eq:4}), (\ref{eq:5}) and (\ref{eq:7}) it results that
$\langle s\rangle\sim \xi^2/P_a$. Moreover, in this models $P_a$ has a finite value
and, therefore, one obtains the mentioned scaling law $\langle s\rangle\sim\xi^2$.

On the contrary, in the model considered here $\langle s\rangle$ can not be related
to $\xi$ using these arguments. For fixed $\epsilon$ from eqs. (\ref{eq:5}) and
(\ref{eq:7}) one obtains that $\langle s\rangle \sim 1/P_a$. Thus, from the energy
balance invoked above we cannot say anything about the scaling of $\langle s\rangle$
with $\xi$ (an exponent 2 will be an accidental coincidence) and, therefore, this
model belongs to a new universality class. 

\section{Numerical simulations and discussion}
\label{sec:numerical}

In this sections results obtained from numerical simulations of the model studied
above are presented. The simulations were performed using $\epsilon=1$, $L=4096$ and
$\zeta$ ranging from $2^0$ to $2^{10}$ ($\epsilon_{eff}=1/\zeta$ ranging from 1 to
$2^{-10}$). For these values the condition $L\ll\xi$ was observed to be satisfied. 
Statistics was taken over $10^8$ avalanches after the system reached the stationary
state. 

\begin{figure}
\centerline{\psfig{file=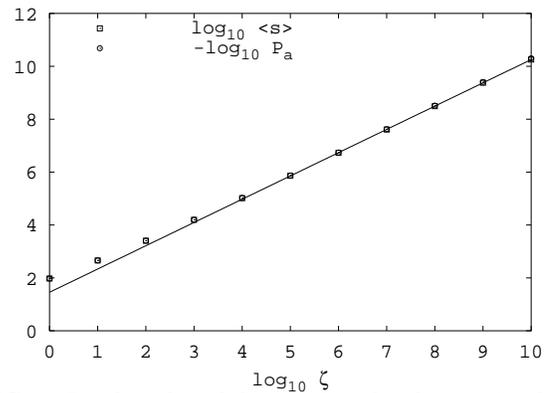,width=3in,angle=-90}}
\caption{Log-log plot of the mean avalanche size (excluding avalanches with size
$s=0$) as a function $\zeta$. It can be clearly seen that it scales as $P_a^{-1}$,
the probability per unit step to obtain an avalanche with $s>0$. The line is a
linear fit to the high $\zeta$ interval.} 
\label{fig:1}
\end{figure}

Before entering in the analysis of the statistics of the avalanches let us check the
validity of eq. (\ref{eq:5}). The log-log plot of $\langle s\rangle$ vs. $\zeta$ is
shown in fig. \ref{fig:1}. A clear linear behavior is observed for
$\log_{10}\zeta\geq5$ suggesting that above this value simple scaling applies. On
top of this points the numerically computed values of $1/P_a$ are plotted obtaining
an overlap in agreement with eq. (\ref{eq:5}). If the scaling relation $\langle
s\rangle \sim\xi^2$ where valid, using eq. (\ref{eq:4}), $\langle
s\rangle\sim\zeta$.  However, a linear fit to this log-log plot gives a slope
$\sim0.9$.

\begin{figure}
\centerline{\psfig{file=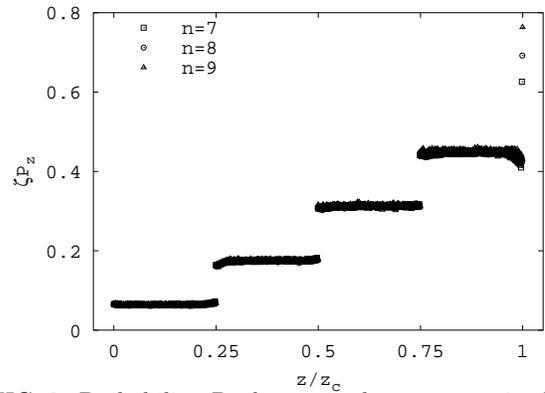,width=3in,angle=-90}}
\caption{Probability $P_z$ that a site has energy $z$ in the stationary state, for
different values of $\zeta=2^n$. $z$ is expressed in units of the threshold
$z_c=4\zeta+1$ while $P_z$ has being rescaled by an amount $\zeta$ because with
increasing $\zeta$ the density of $z/z_c$ values increases as $\zeta$.}
\label{fig:2}
\end{figure}

The fact that this scaling relation thus not hold is clearly shown in fig. 
\ref{fig:2}, where the stationary energy distribution is shown. As it can be seen
$P_a=P_{z_c-1}$ does not decreases as $1/\zeta$ but slower, which explains why
$\langle s \rangle$ increases with $\zeta$ with a slope smaller than 1. The rest of
the distribution scales like $1/\zeta$ which is just a consequence of the
increase of the density of possible values of $z$. 

The avalanche statistics will be characterized by: the number of toppling events $s$
and steps $t$ required to reach an stable configuration, the number of sites $a$
"touched" by the avalanche, and the characteristic radius of the cluster formed by
these sites $r$. The main goal of the simulations is to determine the probability
densities $p_x(x,\zeta)$ ($x=s,t,a,r$) in the stationary state.

One can easily see that $s=a$, in other words sites topples only once within an
avalanche. In this model, as a difference with the original BTW model, only one wave
of topplings takes place. The first wave if generated from an initial site with
height $z=z_c=4\zeta+\epsilon$. When this site topples it transfers an amount equal
to $z_c$ to its nearest neighbors and, therefore, ends with energy $z=0$. The best
we can have to obtain a second toppling at this site is that its four nearest
neighbors also become active. In such a case the initial side will receive
$4\zeta<z_c$ units of energy, which is not enough to make it active again. Hence, no
second wave will be obtained yielding $s=a$.

Since the waves are known to satisfy well defined finite-size scaling properties and
in the present model an avalanche is made by one wave it is expected that the
distributions $p_x(x,\zeta)$ also satisfy a finite-size scaling. However, the
scaling exponents will not necessarily be those obtained for the scaling of waves
because, in the present model, conservation does not introduce any scaling relation
among the different scaling exponents.

If finite size scaling applies then these densities will satisfy
\begin{equation}
p_x(x,\zeta)=x^{-\tau_x}{\cal G}[x/x_c(\zeta)],
\label{eq:8}
\end{equation}
where $\tau_x$ is the power law exponent characterizing the self-similar regime and
$x_c$ is a cut off above which the distribution deviates from a power law and has a
fast decay given by ${\cal G}$. The validity of this scaling form is supported by
the numerical results. The cut off $x_c$ is determined by the existence of the
characteristic length $\xi\sim\zeta^\nu$ and is expected to scale as
$x_c\sim\xi^{D_x}\sim\zeta^{d_x}$, where $d_x=D_x\nu$ is an effective fractal
dimension.

To compute the exponents $\tau_x$ and $d_x$ the moment analysis technique
introduced by De Menech {\em et al} \cite{demenech98} is used. The moments of the
probability density in eq. (\ref{eq:8}) are given by
\begin{equation}
\langle x^q\rangle=\int_0^\infty dx p(x) x^q\sim \zeta^{\sigma_x(q)},
\label{eq:9}
\end{equation}
where
\begin{equation}
\sigma_x(q)=(1-\tau_x)d_x+d_xq.
\label{eq:10}
\end{equation}
The last equivalence in eq. (\ref{eq:9}) is valid for values of $q$ not to small,
for which the precise form of $p_x(x,\zeta)$ at small $x$ is not important. 

\begin{figure}
\centerline{\psfig{file=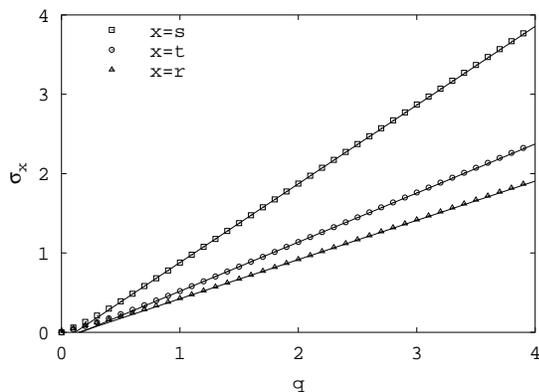,width=3in,angle=-90}}
\caption{Moment exponent $\sigma_x(q)$ for different values of $q$ and $x=s,t,r$.
The lines are linear fits [$\sigma_x(q)=(1-\tau_x)d_x+d_xq$] to the interval $1\geq
q\leq3$. The resulting exponents $\tau_x$ and $d_x$ are shown in tab. \ref{tab:1}.} 
\label{fig:3}
\end{figure}

$\sigma_x(q)$ can be determined from a linear fit to the log-log plot of $\langle
x^q\rangle$ vs. $\zeta$. The resulting values using $\zeta=2^5,\ 2^6,\ldots,\
2^{10}$ are shown in fig. \ref{fig:3}. In all cases ($x=s,t,r$) for $q$ larger than
1 a well defined linear dependence is observed. From the linear fit (see eq.
\ref{eq:10}) to these plots the exponents $\tau_x$ and $d_x$ are computed. The
results are shown in table \ref{tab:1}. 

The exponent $\nu$ is very close to $1/2$ in very good agreement with the scaling
arguments of previous section. On the other hand, $d_s$ is quite close to 1 which
implies that the avalanche size (or area) scale as $s\sim r^2$, i.e. avalanches are
compact $D_s=2$. With this value, the scaling relation $(2-\tau_s)D_s=2$ yields the
power law exponent $\tau_s=1$ which is clearly in disagreement with the value
computed numerically. The reason for this result is that conservation does not
introduce any scaling relation as it generally occurs in sandpile models
\cite{vazquez00}. 

\begin{figure}
\centerline{\psfig{file=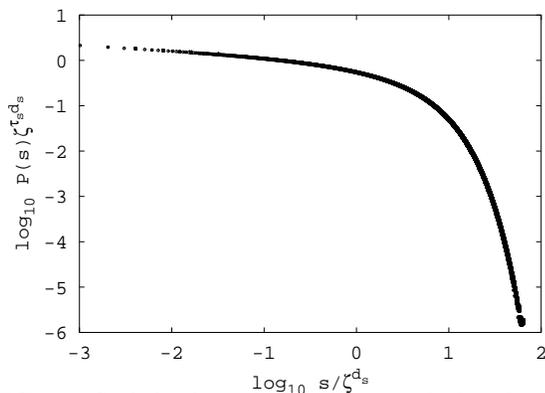,width=3in,angle=-90}}
\caption{Scaled plot of the integrated distribution of avalanche sizes (or area
since $s=a$ in this model) using the exponents displayed in tab \ref{tab:1}.}
\label{fig:4}
\end{figure}

\begin{figure}
\centerline{\psfig{file=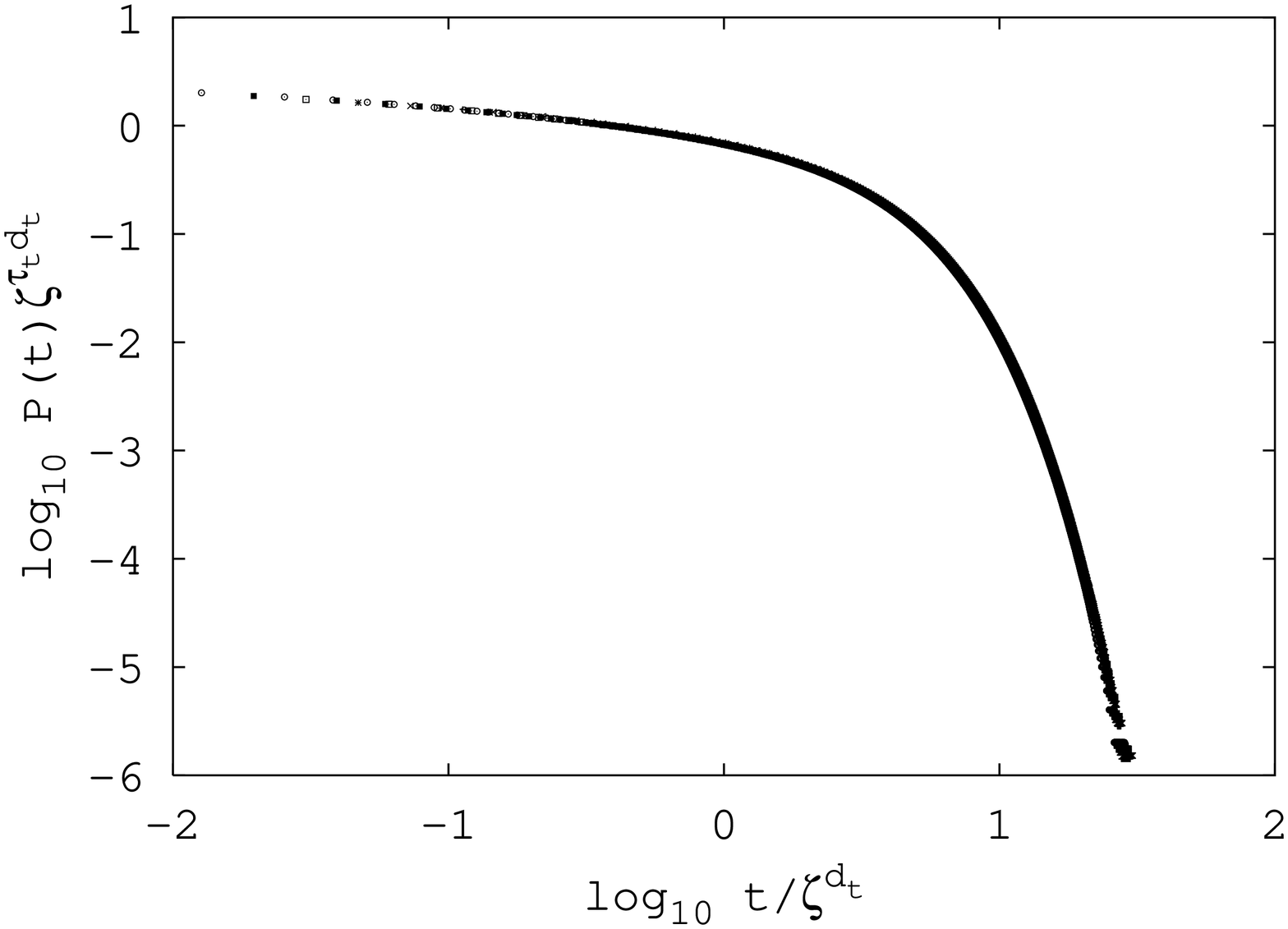,width=3in,angle=-90}}
\caption{Scaled plot of the integrated distribution of avalanche durations using the
exponents displayed in tab \ref{tab:1}.}
\label{fig:5}
\end{figure}

\begin{figure}
\centerline{\psfig{file=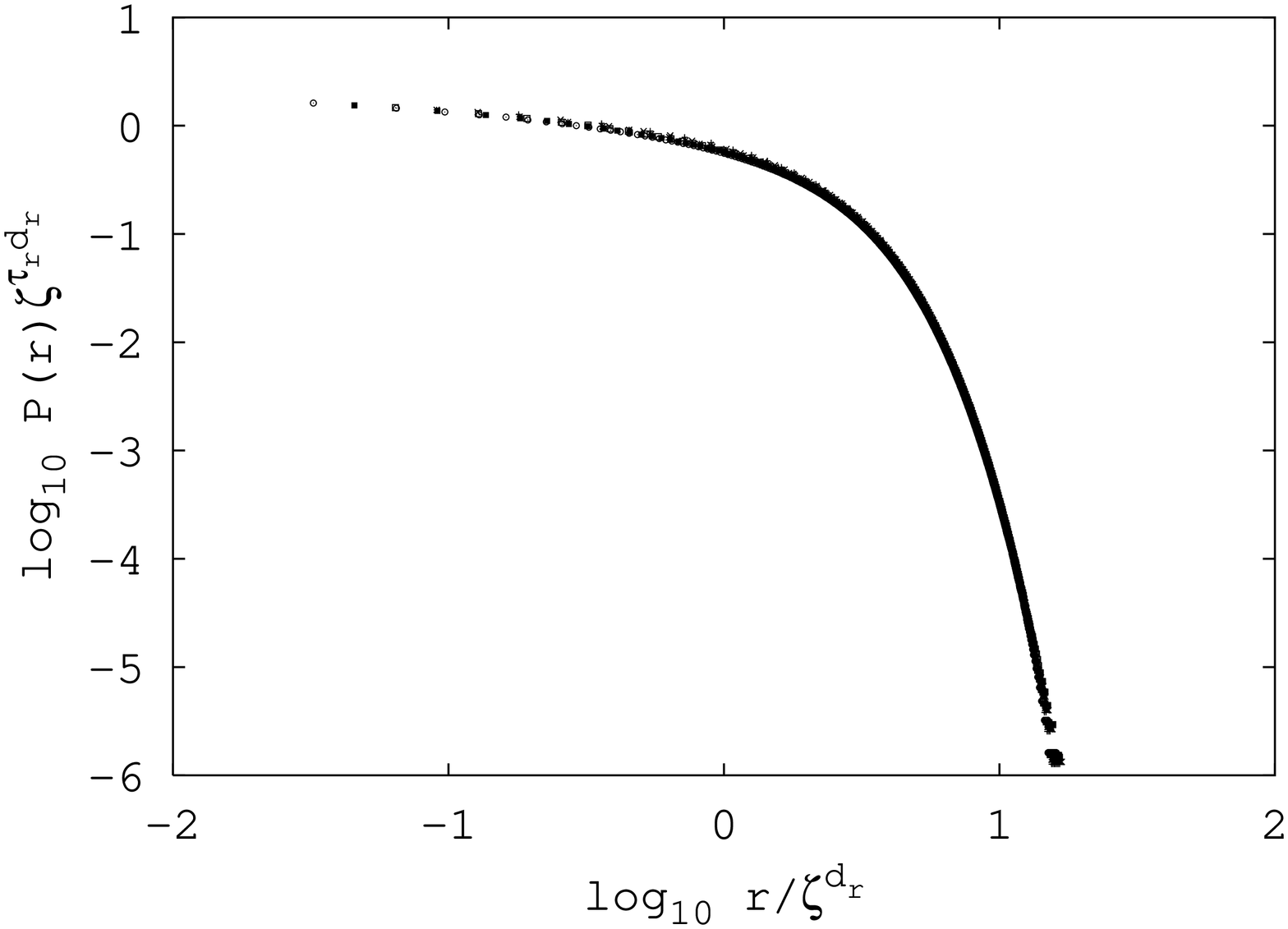,width=3in,angle=-90}}
\caption{Scaled plot of the integrated distribution of avalanche radius using the
exponents displayed in tab \ref{tab:1}.}
\label{fig:6}
\end{figure}

The exponents computed using the moment analysis technique can be checked using
rescaled plots of the integrated distribution $P_x(x,\zeta)=\int_x^\infty
dxp_x(x,\zeta)$. The resulting plots are shown in figs. \ref{fig:4}, \ref{fig:5},
and \ref{fig:6}. The scaling works quite good supporting the validity of the
reported exponents. 

In the literature we can find other sandpile models with local dissipation and BTW
like toppling rule \cite{manna90,ghaffari97,chessa98}. In the models considered in
\cite{manna90} and \cite{ghaffari97} the energy profile is continuous and the
dissipation rate per toppling event $\epsilon$ is a control parameter which can take
any real value and, therefore, can be tuned to zero. Another feature of these models
is that only one wave of toppling can take place and, therefore, for any finite
$\epsilon$ the model is in a different universality class from that of the BTW
model.

On the other hand in \cite{chessa98} the energy profile is discrete as in the
original BTW model at the prize of introducing stochasticity in the model. In this
case with a probability $p$ energy is fully dissipated yielding an average
dissipation per toppling event $\epsilon=2dp$. Clearly $p$ may take any real
variable between 0 and 1 and, therefore, also in this case the dissipation per
toppling event can be fine tuned to zero. As a difference with the models described
in the previous paragraph, in this case multiple toppling of a site within an
avalanche is possible, which make it closer to the original BTW model. Moreover, the
use of finite size scaling techniques can be also questioned and a multi-fractal
analysis may be more appropriate \cite{vespignani00}, which is another
characteristic feature of the BTW model \cite{tebaldi99}. All this elements together
with the numerical results reported in \cite{chessa98} suggest that in the limit
$p\rightarrow1$ ($\epsilon\rightarrow0$) this model belong to the same universality
class of the BTW model.

A common feature of all this models \cite{manna90,ghaffari97,chessa98} is that
$\langle s\rangle\sim\epsilon^{-1}$ as predicted by the field theory approach of
Vespignani {\em et al} \cite{vespignani97}, leading to the scaling relation
$(2-\tau_s)D_s=2$. On the contrary, in the present model the scaling of $\langle
s\rangle$ with $\epsilon_{eff}$ is not known and conservation does not introduce the
above scaling relation. Hence, the model introduced by Tsuchiya and Katori belongs
to different class among sandpile models.

\section{Summary and conclusions}

A non conservative Abelian sandpile model with a BTW like toppling rule has been
studied. The model can be though as the only possible generalization of the BTW
model to include local dissipation without introducing stochasticity in the toppling
rule and keeping a discrete energy profile. However, the scaling approach and
numerical the simulations reported here show that it does not belong to the
universality class of the BTW model, not even to the universality class of any
sandpile model previously considered in the literature. 

\section*{Acknowledgements}

I thanks R. Pastor Satorras and A. Vespignani for useful comments and discussion.
The numerical simulations where performed using the computing facilities at the
ICTP.

\begin{table}
\begin{tabular}{llllll}
$d_s$ & $d_t$ & $d_r=\nu$ & $\tau_s$ & $\tau_t$ & $\tau_r$\\
\hline
0.994(5) & 0.630(5) & 0.495(5) & 1.11(1) & 1.16(1) & 1.14(1)\\
\hline
$D_s=d_s/\nu$ & $z=D_t=d_t/\nu$ & & & &\\
\hline
2.01(1) & 1.27(1) 
\end{tabular}
\caption{Scaling exponents obtained from linear fits 
[$\sigma_x(q)=(1-\tau_x)d_x+d_xq$] to the data shown in fig. \ref{fig:3}.}
\label{tab:1}
\end{table}

\end{multicols}

\end{document}